\documentclass[a4paper,english,11pt,amssymb,nofootinbib,superscriptaddress]{revtex4-2}
\usepackage{lmodern}
\usepackage{lmodern}
\usepackage[T1]{fontenc}
\usepackage[latin9]{inputenc}
\usepackage{color}
\usepackage{babel}
\usepackage{verbatim}
\usepackage{float}
\usepackage{amsmath}
\usepackage{amssymb}
\usepackage{graphicx}
\usepackage{esint}
\usepackage[unicode=true,pdfusetitle,
 bookmarks=true,bookmarksnumbered=false,bookmarksopen=false,
 breaklinks=false,pdfborder={0 0 1},backref=false,colorlinks=false]
 {hyperref}

\makeatletter

\pdfpageheight\paperheight
\pdfpagewidth\paperwidth

\usepackage{babel}

\pdfpageheight\paperheight
\pdfpagewidth\paperwidth

\@ifundefined{textcolor}{}{%
 \definecolor{BLACK}{gray}{0}
 \definecolor{WHITE}{gray}{1}
 \definecolor{RED}{rgb}{1,0,0}
 \definecolor{GREEN}{rgb}{0,1,0}
 \definecolor{BLUE}{rgb}{0,0,1}
 \definecolor{CYAN}{cmyk}{1,0,0,0}
 \definecolor{MAGENTA}{cmyk}{0,1,0,0}
 \definecolor{YELLOW}{cmyk}{0,0,1,0}
}

\@ifundefined{definecolor}{color}{}
\@ifundefined{definecolor}{color}{}

\@ifundefined{definecolor}{color}{}

\pdfpageheight\paperheight
\pdfpagewidth\paperwidth

\@ifundefined{textcolor}{}{%
 \definecolor{BLACK}{gray}{0}
 \definecolor{WHITE}{gray}{1}
 \definecolor{RED}{rgb}{1,0,0}
 \definecolor{GREEN}{rgb}{0,1,0}
 \definecolor{BLUE}{rgb}{0,0,1}
 \definecolor{CYAN}{cmyk}{1,0,0,0}
 \definecolor{MAGENTA}{cmyk}{0,1,0,0}
 \definecolor{YELLOW}{cmyk}{0,0,1,0}
 }

\@ifundefined{definecolor}{color}{}
\usepackage{latexsym}\usepackage{bm}

\makeatother

\begin{document}
\title{Israel-Wilson-Perjes Metrics in a Theory with a Dilaton Field }
\author{Metin G\"{u}rses}
\email{gurses@fen.bilkent.edu.tr}

\affiliation{{\small{}Department of Mathematics, Faculty of Sciences}~\\
{\small{}Bilkent University, 06800 Ankara,} T\"{u}rkiye}
\author{Tahsin \c{C}a\u{g}r\i{} \c{S}i\c{s}man}
\email{tahsin.c.sisman@gmail.com}

\affiliation{Department of Astronautical Engineering,~\\
 University of Turkish Aeronautical Association, 06790 Ankara, T\"{u}rkiye}
\author{Bayram Tekin}
\email{btekin@metu.edu.tr}

\affiliation{Department of Physics, Middle East Technical University, 06800 Ankara,
T\"{u}rkiye}
\begin{abstract}
We are interested in the charged dust solutions of the Einstein field
equations in stationary and axially symmetric spacetimes; and inquire
if the naked singularities of the Israel-Wilson-Perjes (IWP) metrics
can be removed. The answer is negative in four dimensions. We examine
whether this negative result can be avoided by adding scalar or dilaton
fields. We show that IWP metrics also arise as solutions of the Einstein-Maxwell
system with a stealth dilaton field. We determine the IWP metrics
completely in terms of one complex function satisfying the Laplace
equation. With the inclusion of the stealth dilaton field, it is now
possible to add a perfect fluid source. In this case the field equations
reduce to a complex cubic equation. Hence this procedure provides
interior solutions to each IWP metric; and it is possible to cover
all naked singularities inside a compact surface where there is matter
distribution. 
\end{abstract}
\maketitle

\section{Introduction}

In four dimensions, conformastatic \textcolor{black}{or conformastat
metrics \citep{step},} 
\begin{equation}
ds^{2}=-\lambda^{-2}\left(\vec{x}\right)\,dt^{2}+\lambda^{2}\left(\vec{x}\right)d\vec{x}\cdot d\vec{x},
\end{equation}
solve the Einstein-Maxwell-dust field equations with the vector potential
$A^{\mu}=\left(2\lambda\left(\vec{x}\right),0,0,0\right)$ if the
metric function $\lambda\left(\vec{x}\right)$ satisfies 
\begin{equation}
\nabla^{2}\lambda+\frac{1}{2}\,\rho_{m}\lambda^{3}=0,\label{eq:MP_electrostatic_potential_eqn}
\end{equation}
where $\nabla^{2}$ denotes the three-dimensional Laplace operator
in flat Cartesian coordinates; and $\rho_{m}$ is the mass density
of the dust distribution with the four-velocity $u_{\mu}=\delta_{\mu}^{t}/\lambda\left(\vec{x}\right)$
\citep{gur1,gur2}. The electric charge density $\rho_{e}$ and the
mass density $\rho_{m}$ are equal; hence, the system is ``extremely''
charged. The $\rho_{m}=0$ case, i.e. the conformastatic solutions
of the Einstein-Maxwell equations, the so called Majumdar-Papapetrou
(MP) metrics \citep{step,maj,pap}, represent gravitational fields
of multiple extremely charged black holes \citep{hh} of electrovacuum.
One can extend these solutions by adding a charged dust distribution
\citep{gur1,gur2,gur3} (see also the references therein) where several
interesting thin shell charged dust solutions without the singularities
of the MP metrics were obtained.

Extension of the static MP solutions to the stationary case was done
in \citep{iw,per}, and a relevant question is to generalize the conformastatic
MP metrics with extremely charged dust to conformastationary Israel-Wilson-Perjes
(IWP) metrics with charged dust. This problem, in a sense, asks one
to search for dust sources for the IWP metrics. Here we show that
conformastationary spacetimes do not support charged dust solutions
\citep{chr1}. The integrability conditions of the rotation velocity
and the magnetic potential vectors reduce the problem either to the
sourceless case, i.e., the IWP metrics \citep{iw,per} or to MP metrics
with dust \citep{gur1,gur2,gur3}.

To summarize our physical motivations and our findings in this work, let us note tha following:  Hartle and Hawking  wrote the influential paper \citep{hh}, in which, among other things related to the stationary solutions, in their own words they claim the following:  {\it ``We also analyse some of stationary solutions of the Einstein-Maxwell equations discovered by Israel and Wilson. If space is asymptotically Euclidean we find that all of
these solutions have naked singularities."} As the nature of the singularity in gravity is an extremely important issue, this work received a lot of attention. There are in principle solutions of Einstein-Maxwell theory in which naked singularities are unavoidable. What we will show below is to answer the the following question: by adding a source to the Einstein-Maxwell system, can one avoid these naked singularities ? There are two non-trivial facets to this problem: first of all, the full theory with the source must admit the Israel-Wilson-Perjes metrics as solutions; and secondly the naked singularities must be avoided. Both of these questions have, by no means obvious, answers. In fact, our first attempt to add a perfect fluid source (with pressure and density) will yield field equations that do not support these solutions with a non-zero pressure. However, the field equations force one to consider the dust source (i.e. a perfect fluid with zero pressure). So, with the inclusion of a dust, Einstein-Maxwell system still supports these solutions, which is an answer to the first question. But, unfortunately, the introduced dust does not remove the naked singularity.  So, this is our theorem (see the theorem below): {\it the IWP metrics with charged perfect source reduce to either Majumdar Papapetrau metrics (the no rotation case) or to the IWP metrics without a source.} This theorem by itself  an important contribution to general relativity. Hence, the Hartle-Hawking no-go result is extended to the case of Einstein-Maxwell-dust theory in our work. We then  answer the second question, that is the question of removing the singularities, with the addition of a scalar field. It really is remarkable that such a complicated system both allows the  IWP metrics as solutions; and removes their stubborn  naked singularities.

The layout of the paper is as follows: In Sec\@.~II, we study the
conformostationary spacetimes as solutions of Einstein-Maxwell field
equations with a perfect fluid distribution, and show that such a
configuration is not possible. In Sec.~III, we introduce the dilaton
field to the previous system, and we show that in four dimensions,
IWP metrics with dust source is possible if the dilaton field has
a vanishing energy-momentum tensor (that is, the dilaton becomes a
stealth field). In Sec.~IV, we reduce the field equations to a complex
potential equation. In Sec.~V, we discuss the non-vacuum cases. In
the Appendix, we give an action formulation of the field equations.

\section{Conformostationary Spacetimes}

Einstein-Maxwell-perfect fluid field equations are given as 
\begin{align}
G_{\mu\nu} & =\frac{1}{2}T_{\mu\nu}+\left(\rho_{m}+p\right)u_{\mu}u_{\nu}+pg_{\mu\nu},\label{eq:EoM_grav_Einstein-Maxwell}\\
\nabla_{\alpha}\,F^{\alpha\mu} & =\rho_{e}u^{\mu},\label{eq:EoM_electromag_Einstein-Maxwell}
\end{align}
where the Einstein tensor $G_{\mu\nu}$ is defined as $G_{\mu\nu}\equiv R_{\mu\nu}-\frac{1}{2}g_{\mu\nu}R$,
$\rho_{m}$ and $\rho_{e}$ are the mass and electric charge densities;
$F_{\mu\nu}=\partial_{\mu}\,A_{\nu}-\partial_{\nu}\,A_{\mu}$ with
$A_{\mu}=\left(A_{0},\vec{A}\right)$; and the energy-momentum tensor
of the Maxwell field reads 
\begin{equation}
T_{\mu\nu}=F_{\mu\alpha}\,F_{\nu}\,^{\alpha}-\frac{1}{4}g_{\mu\nu}F^{2},\label{eq:T^m}
\end{equation}
with the definition $F^{2}=F_{\sigma\rho}\,F^{\sigma\rho}$. Taking
the metric as 
\begin{equation}
ds^{2}=-f^{2}\left(dt+\vec{\Omega}\cdot d\vec{x}\right)^{2}+\frac{1}{f^{2}}\,d\vec{x}\cdot d\vec{x},\label{met}
\end{equation}
where the metric functions $f$ and $\vec{\Omega}$ depend on the
spatial coordinates $\vec{x}$, and one has $u_{\mu}=f\left(1,\vec{\Omega}\right)$.
The spatial indices are raised and lowered with the Kronecker-delta
$\delta_{ij}$. The field equations (\ref{eq:EoM_grav_Einstein-Maxwell})
and (\ref{eq:EoM_electromag_Einstein-Maxwell}) reduce to 
\begin{eqnarray}
 &  & \partial_{i}\Omega_{j}-\partial_{j}\Omega_{i}=2\epsilon_{ijk}\,\Im(\lambda\partial_{k}\bar{\lambda}),\label{omega1}\\
 &  & \partial_{i}A_{j}-\partial_{j}A_{i}=\frac{1}{f^{2}}\,\epsilon_{ijk}\,\partial_{k}\chi+\Omega_{j}\,\partial_{i}A_{0}-\Omega_{i}\,\partial_{j}A_{0},\label{omega2}
\end{eqnarray}
where $\lambda$ is a complex function, with real and imaginary parts
given as $\Re(\lambda)$, $\Im(\lambda)$, while $|\lambda|$ is its
magnitude; and in terms of $\lambda$, the real functions $f$, $A_{0}$,
and $\chi$ are given as 
\begin{eqnarray}
\chi=\frac{2\Im(\lambda)}{|\lambda|^{2}},\qquad f=\frac{1}{|\lambda|},\qquad A_{0}=-\frac{2\Re(\lambda)}{|\lambda|^{2}}.\label{met}
\end{eqnarray}
Then the remaining field equations give vanishing pressure $p=0$,
the mass $\rho_{m}$ and charge $\rho_{e}$ densities are determined
via 
\begin{eqnarray}
 &  & \Re(\lambda\nabla^{2}\bar{\lambda})+\frac{1}{2}\rho_{m}\,|\lambda|^{4}=0,\label{density1}\\
 &  & \Re(\lambda^{2}\nabla^{2}\bar{\lambda})\,+\frac{1}{2}\rho_{e}|\lambda|^{5}=0.\label{density2}
\end{eqnarray}
Defining the vector potential as $\vec{A}=\vec{B}+\vec{\Omega}\,A_{0}$,
one finds 
\begin{equation}
\partial_{i}B_{j}-\partial_{j}B_{i}=-2\,\epsilon_{ijk}\partial_{k}\Im(\lambda).\label{bb1}
\end{equation}
We have the following Lemma.

\noindent \textbf{Lemma}: \textit{Integrability of (\ref{omega1})
and (\ref{bb1}) imply that 
\begin{eqnarray}
 &  & \Im(\lambda\nabla^{2}\bar{\lambda})=0,\label{int1}\\
 &  & \nabla^{2}\,\Im(\lambda)=0.\label{int2}
\end{eqnarray}
} Proof of the Lemma is easier when we convert the equations (\ref{omega1})
and (\ref{bb1}) to the following forms 
\begin{eqnarray}
 &  & \epsilon^{ijk}\,\partial_{i}\Omega_{j}=2\,\Im(\lambda\partial_{k}\bar{\lambda}),\\
 &  & \epsilon^{ijk}\,\partial_{i}B_{j}=2\partial_{k}\Im(\lambda).
\end{eqnarray}
Taking the divergence of both sides of these equations, we find (\ref{int1})
and (\ref{int2}). Integrability condition (\ref{int1}) implies,
from (\ref{density1}), 
\begin{equation}
\nabla^{2}\lambda+\frac{1}{2}\rho_{m}\lambda^{2}\bar{\lambda}=0\,,\label{lap}
\end{equation}
and the integrability condition (\ref{int2}) implies, from (\ref{density2}),
\begin{equation}
\Re(\lambda)\rho_{m}=|\lambda|\rho_{e}.
\end{equation}

\noindent \vspace{0.3cm}
 Then we have the following theorem.

\noindent \vspace{0.3cm}
 \textbf{Theorem}: \textit{Conformostationary metrics (\ref{met})
do not support Einstein-Maxwell-dust field equations. They reduce
to either Israel-Wilson-Perjes i.e., $\rho_{m}=\rho_{e}=0$ or to
Majumdar-Papapetrou metrics i.e., $\vec{\Omega}=0$.}

\noindent \vspace{0.3cm}
 Proof of the theorem comes from the integrability conditions (\ref{int1})
and (\ref{int2}) and hence from (\ref{lap}), one has
\begin{equation}
\nabla^{2}\Im(\lambda)+\frac{1}{2}\rho_{m}\,|\lambda|^{2}\Im(\lambda)=0.\label{denk1}
\end{equation}
Since the first term on the left-hand side of (\ref{denk1}) vanishes
due to (\ref{int2}), then the second term gives $\rho_{m}\,\Im{(\lambda)}=0$.
Hence either $\rho_{m}=0$ (Israel-Wilson-Perjes spacetimes) or $\lambda=\bar{\lambda}$
leading to, without losing any generality, $\vec{\Omega}=0$ which
corresponds to the Majumdar-Papapetrou spacetimes and $\lambda$ satisfies
the (\ref{eq:MP_electrostatic_potential_eqn}).

\noindent \vspace{0.3cm}
 \textbf{Remark}: Our theorem is consistent with the results of \citep{chr1}.
In \citep{chr1} the authors consider also a magnetic current. In
the absence of the magnetic current density, then their result reduces
to our result.

\section{Inclusion of a Dilaton Field}

\noindent Following \citep{gur-sar1} and \citep{gur-sar2} we take
the metric as $g_{\mu\nu}=e^{\frac{2\phi}{2-D}}\,\left(-u_{\mu}\,u_{\nu}+h_{\mu\nu}\right)$
in $D$ dimensions, where $u_{\mu}=e^{\phi}\left(1,\vec{q}\right)$
and $h_{\mu\nu}$ is a constant two tensor with $u^{\mu}\,h_{\mu\nu}=0$.
Taking $F_{\alpha\beta}=\partial_{\alpha}\,u_{\beta}-\partial_{\beta}\,u_{\alpha}$,
one finds 
\begin{eqnarray}
 &  & G_{\mu\nu}=\frac{4-D}{2(2-D)}\,T_{\mu\nu}^{\phi}+\frac{1}{2}\,e^{\frac{2\phi}{2-D}}\,T_{\mu\nu}^{M}+\left(\rho_{m}+p\right)\,v_{\mu}\,v_{\nu}+p\,g_{\mu\nu},\label{denk1-1}\\
 &  & \nabla_{\alpha}\,\left(\,e^{\frac{2\phi}{2-D}}\,F^{\alpha}\,_{\nu}\right)-\frac{1}{4}\,e^{\frac{4\phi}{2-D}}\,F^{2}\,u_{\nu}=\left(\rho_{e}+p\right)\,\alpha^{2}\,u_{\nu},\label{denk2}\\
 &  & \square\phi+\frac{1}{4}\,e^{\frac{2\phi}{2-D}}\,F^{2}=\rho_{e}-p,\label{denk3}
\end{eqnarray}
where $v_{\mu}=\alpha u_{\mu}$ with $\alpha=e^{\frac{\phi}{2-D}}$,
we have $g^{\mu\nu}\,u_{\mu}\,u_{\nu}=-1/\alpha^{2}$ and $g^{\mu\nu}\,v_{\mu}\,v_{\nu}=-1$.
In addition, the energy-momentum tensor of the dilaton field is defined
as 
\begin{equation}
T_{\mu\nu}^{\phi}=\partial_{\mu}\phi\partial_{\nu}\phi-\frac{1}{2}g_{\mu\nu}\partial_{\rho}\phi\partial^{\rho}\phi,
\end{equation}
while the energy-momentum tensor for the electromagnetic field $T_{\mu\nu}^{M}$
is given in (\ref{eq:T^m}).

\vspace{0.5cm}
 For $D=4$ we have, 
\[
g_{\mu\nu}=e^{-\phi}\,\left(-u_{\mu}\,u_{\nu}+h_{\mu\nu}\right),
\]
and the gravitational field equation takes a special form for which
the backreaction of the scalar field disappears as the energy momentum
tensor of the dilaton field drops from the gravitational field equation,
and $\phi$ becomes a stealth dilaton field;

\noindent 
\begin{eqnarray}
 &  & G_{\mu\nu}=\frac{1}{2}\,e^{-\phi}\,T_{\mu\nu}^{M}+\left(\rho+p\right)\,v_{\mu}\,v_{\nu}+p\,g_{\mu\nu},\\
 &  & \nabla_{\alpha}\,\left(\,e^{-\phi}\,F^{\alpha}\,_{\nu}\right)-\frac{1}{4}\,e^{-2\phi}\,F^{2}\,u_{\nu}=\left(\rho+p\right)e^{-\phi}\,u_{\nu},\\
 &  & \square\phi+\frac{1}{4}\,e^{-\phi}\,F^{2}=\rho-p.
\end{eqnarray}
We find that $p=0$, then

\begin{eqnarray}
 &  & G_{\mu\nu}=\frac{1}{2}\,e^{-\phi}\,T_{\mu\nu}^{M}+\rho\,v_{\mu}\,v_{\nu},\label{eq:EoM_gravity_p=00003D0}\\
 &  & \nabla_{\alpha}\,\left(\,e^{-\phi}\,F^{\alpha}\,_{\nu}\right)-\frac{1}{4}\,e^{-2\phi}\,F^{2}\,u_{\nu}=\rho e^{-\phi}\,u_{\nu},\label{eq:EoM_electromagnetic_p=00003D0}\\
 &  & \square\phi+\frac{1}{4}\,e^{-\phi}\,F^{2}=\rho.\label{eq:EoM_dilaton_p=00003D0}
\end{eqnarray}

The last equation is also the zeroth component of the first Maxwell's
equations. The spatial components of Maxwell's equations (\ref{eq:EoM_electromagnetic_p=00003D0})
reduce to 
\begin{equation}
\partial_{i}\,\left(e^{-2\phi}\,F_{ij}\right)=0.
\end{equation}
The dilaton equation (\ref{eq:EoM_dilaton_p=00003D0}) is not independent,
and can be obtained by contracting $u^{\nu}$ with the Maxwell's equation
(\ref{eq:EoM_electromagnetic_p=00003D0}). The matter density $\rho_{m}$
needs to be equal to the charge density $\rho_{e}$ as $\rho=\rho_{m}=\rho_{e}$.

\section{Simplified Field Equations}

Maxwell and dilaton field equations take the form 
\begin{eqnarray}
 &  & \partial_{k}\,\left(e^{\phi}\,f_{ik}\right)=0,\label{max}\\
 &  & \nabla^{2}\phi-\frac{1}{2}\,\vec{\nabla}\phi\cdot\vec{\nabla}\phi+\frac{1}{2}\,e^{2\phi}\,f_{ik}\,f_{ik}=\rho\,e^{-\phi},
\end{eqnarray}
where $f_{ij}=\partial_{i}\,q_{j}-\partial_{j}\,q_{i}$. Maxwell's
equations (\ref{max}) imply that

\begin{equation}
f_{ij}=e^{-\phi}\,\varepsilon_{ijk}\,\partial_{k}\,\chi,
\end{equation}
where $\chi$ is a function satisfying $\vec{\nabla}\cdot\left(e^{-\phi}\vec{\nabla}\chi\right)=0$.
Then

\begin{equation}
\nabla^{2}\phi-\frac{1}{2}\,\vec{\nabla}\phi\cdot\vec{\nabla}\phi+\frac{1}{2}\vec{\nabla}\chi\cdot\vec{\nabla}\chi=e^{-\phi}\,\rho.
\end{equation}

\vspace{0.5cm}
 \textbf{Electro-vacuum Case}: For $\rho=0$ we have 
\begin{equation}
f_{ij}=e^{-\phi}\,\varepsilon_{ijk}\,\partial_{k}\,\chi,
\end{equation}
where $\chi$ is a function satisfying the equation, $\vec{\nabla}\cdot\left(e^{-\phi}\vec{\nabla}\chi\right)=0$.
Then 
\begin{equation}
\nabla^{2}\phi-\frac{1}{2}\,\vec{\nabla}\phi\cdot\vec{\nabla}\phi+\frac{1}{2}\,\vec{\nabla}\chi\cdot\vec{\nabla}\chi=0.
\end{equation}
We obtain the solutions of the field equation by solving the coupled
nonlinear differential equations:
\begin{eqnarray}
 &  & \vec{\nabla}\cdot\left(e^{-\phi}\vec{\nabla}\chi\right)=0,\label{eqn1}\\
 &  & \nabla^{2}\phi-\frac{1}{2}\,\vec{\nabla}\phi\cdot\vec{\nabla}\phi+\frac{1}{2}\,\vec{\nabla}\chi\cdot\vec{\nabla}\chi=0.\label{eqn2}
\end{eqnarray}
The metric is 
\begin{equation}
ds^{2}=-e^{\phi}\left(dt+\vec{q}\cdot d\vec{x}\right)^{2}+e^{-\phi}d\vec{x}\cdot d\vec{x}.
\end{equation}
Equations (\ref{eqn1}) and (\ref{eqn2}) can be combined as one complex
equation by defining $\xi=\phi\mp i\chi$ 
\begin{equation}
\nabla^{2}\xi-\frac{1}{2}\vec{\nabla}\xi\cdot\vec{\nabla}\xi=0,
\end{equation}
or $\nabla^{2}e^{-\frac{1}{2}\xi}=0$. Hence $e^{-\frac{1}{2}\xi}=\psi_{1}+i\psi_{2}$
where both $\psi_{1}$ and $\psi_{2}$ satisfy the Laplace equation.
We find that 
\begin{equation}
e^{-\phi}=\psi_{1}^{2}+\psi_{2}^{2},~~~~\chi=\pm2\,\tan^{-1}\,\left(\frac{\psi_{2}}{\psi_{1}}\right),
\end{equation}
and 
\begin{equation}
f_{ij}=\partial_{i}\,q_{j}-\partial_{j}\,q_{i}=\varepsilon_{ijk}\,(\psi_{1}\,\partial_{k}\,\psi_{2}-\psi_{2}\,\partial_{k}\,\psi_{1}).
\end{equation}
The solution of the above equation for $\vec{q}$ is given by 
\begin{equation}
\vec{q}=\vec{x}\times\vec{E},\label{rot}
\end{equation}
where \citep{flan} 
\begin{equation}
\vec{E}=\int_{0}^{1}\,\left[\psi_{1}\left(s\vec{x}\right)\,\vec{\nabla}\,\psi_{2}\left(s\vec{x}\right)-\psi_{2}\left(s\vec{x}\right)\,\vec{\nabla}\,\psi_{1}\left(s\vec{x}\right)\right]\,sds.
\end{equation}
Hence the metric is completely determined%
:

\noindent 
\begin{equation}
ds^{2}=-\frac{1}{\psi_{1}^{2}+\psi_{2}^{2}}\,\Big(dt-\vec{E}\cdot(\vec{x}\times d\vec{x})\Big)^{2}+\left(\psi_{1}^{2}+\psi_{2}^{2}\right)\,d\vec{x}\cdot d\vec{x}.\label{met1}
\end{equation}

\section{Non-Vacuum case}

The metric (\ref{met1}) may have singularities without a horizon
enclosing them. To avoid such situations consider the case with $\rho\ne0$.
In this case the field equations become 
\begin{equation}
\nabla^{2}\xi-\frac{1}{2}\vec{\nabla}\xi\cdot\vec{\nabla}\xi=e^{-\phi}\,\rho,
\end{equation}
where $\xi=\phi\pm i\chi$. The above equation can be converted to
\begin{equation}
\nabla^{2}\Lambda+\frac{1}{2}\,\rho\left(\Lambda\bar{\Lambda}\right)\Lambda=0,\label{cubic}
\end{equation}
where $\Lambda=e^{-\frac{1}{2}\xi}$. The vector $\vec{q}$ remains
the same as in the vacuum case (\ref{rot}). For the thin shell model
and other non-vacuum cases it is better to have a different and useful
parametrization other than $\Lambda=e^{-\frac{1}{2}\xi}$ used in
Sec.~IV . Let $\Lambda=Re^{i\theta}$, then and one has 
\[
R=e^{-\phi/2},\qquad\theta=\pm\frac{\chi}{2}.
\]
With this parametrization (\ref{cubic}) reduces to two real equations
as 
\begin{equation}
\nabla^{2}R-\vec{\nabla}\theta\cdot\vec{\nabla}\theta+\frac{1}{2}\rho R^{3}=0,~~~\vec{\nabla}\cdot\left(R^{2}\vec{\nabla}\theta\right)=0.
\end{equation}

We have the following interesting cases:

\textbf{(1)}. When $\theta=$ constant then the IWP metrics reduce
to the MP metrics, but when $R=$ constant, the source density is
$\rho=\frac{2}{R^{2}}\,\vec{\nabla}\theta\cdot\vec{\nabla}\theta$
and the resulting metric is given by

\noindent 
\begin{equation}
ds^{2}=-\frac{1}{R^{2}}\,\Big(dt-\vec{E}\cdot(\vec{x}\times d\vec{x})\Big)^{2}+R^{2}\,d\vec{x}\cdot d\vec{x},
\end{equation}
with 
\begin{equation}
\vec{E}=R^{2}\,\vec{\nabla}\,\int_{0}^{1}\,\theta(s\vec{x})\,sds,
\end{equation}
where $\nabla^{2}\theta=0$. The spacetime obtained is a dust filled
rotating universe.

{

\textbf{(2)}. When $\rho=\rho_{0}\,\theta(F)$, where $\rho_{0}$
is the density function in a region $F>0$. Here $F=0$ defines a
compact surface outside of which is the vacuum (IWP) metric (\ref{met1}).
Hiding the singularities of the IWP metrics inside the compact surface
$F=0$, we get rotating regular charged solutions of
Einstein field equations with a dilaton field.}

\textbf{(3)}. For $\rho=\rho_{0}\,\delta(F)$ where $\rho_{0}$ is
the density function on the thin shell $F=0$, we have a shell model;
in both sides of the thin shell the spacetime is described by the
IWP metric (\ref{met1}) with $\rho=0$. As an example of such a model
let $\rho(x,y,z)=\rho_{0}(x,y)\,\delta(z)$. Above the thin shell
($z>0$) the metric is an IWP metric with the metric functions $(R_{1},\theta_{1})$
and below the thin shell ($z<0$) the metric is also an IWP metric
with the metric functions $(R_{2},\theta_{2})$ (see Figure 1). These
metric functions are continuous at $z=0$ but $\frac{\partial R}{\partial z}$
is not continuous and satisfies the following jump condition: 
\begin{equation}
\frac{\partial R_{1}}{\partial z}-\frac{\partial R_{2}}{\partial z}+\frac{1}{2}\,R^{3}\,\rho_{0}(x,y)=0,~~~~~~\mbox{at}~~~z=0.
\end{equation}
\begin{figure}[H]
\begin{centering}
\includegraphics[scale=0.9]{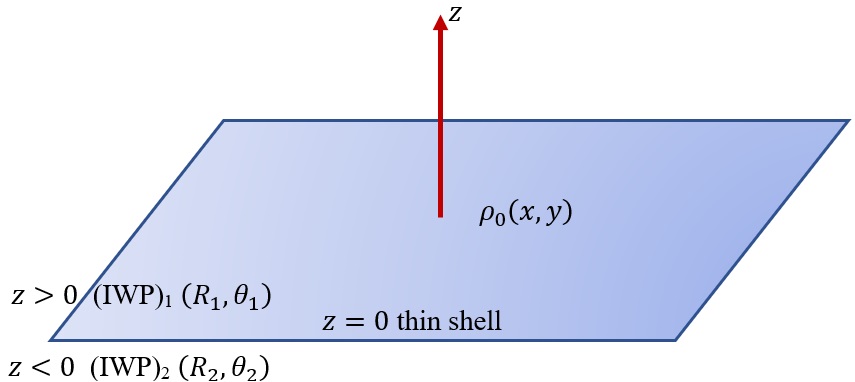} 
\par\end{centering}
\caption{The thin shell is located at $z=0$ as an infinite plane on which
the charge density is $\rho_{0}\left(x,y\right)$. The thin shell
separates two different IWP spacetimes described by the metric functions
$\left(R_{1},\theta_{1}\right)$ and $\left(R_{2},\theta_{2}\right)$.}
\end{figure}

\section{Conclusions and Discussions}

We have considered the IWP metrics in four dimensions and showed that
these metrics have many interesting properties which can be summarized as follows:

\noindent\vskip 0.5 cm 
 (1) We showed that the IWP metrics, with $F_{\mu\nu}=\partial_{\mu}\,A_{\nu}-\partial_{\nu}\,A_{\mu}$
where $A_{\mu}=(A_{0},\vec{A})$, solve the Einstein Maxwell field
equations, but do not admit charged perfect fluids as sources. Furthermore
these metrics contain naked singularities; hence do not represent
the gravitational fields of black holes \citep{hh}.

\noindent\vskip 0.5 cm 
 (2) We showed that the IWP metrics in four dimensions, $g_{\mu\nu}=e^{-\phi}\,\left(-u_{\mu}\,u_{\nu}+h_{\mu\nu}\right)$
or 
\begin{equation}
ds^{2}=-e^{\phi}\,(dt+\vec{q}\cdot d\vec{x})^{2}+e^{-\phi}\,d\vec{x}\cdot d\vec{x},
\end{equation}
for $F_{\mu\nu}=\partial_{\mu}\,u_{\nu}-\partial_{\nu}\,u_{\mu}$,
admit charged dust distributions as sources with a stealth (having a vanishing
energy-momentum tensor) dilaton field $\phi$.

\noindent \vskip 0.5 cm 
(3) For $\rho=0$, we have given the complete solution in terms of
two harmonic functions $\psi_{1}$ and $\psi_{2}$. For $\rho\ne0$
the field equations reduce to a cubic equation for a complex function
$\Lambda$. It is now possible to cover all naked singularities inside
a compact surface $F=0$ where the matter density $\rho\ne0$. Outside
of the compact surface $F=0$ where we let matter distribution the
geometry is described by the IWP metrics with $\rho=0$.

\noindent \vskip 0.5 cm 

 (4) It is possible to obtain exact solutions for multi-IWP universes
separated by thin shells represented with the matter density
\begin{equation}
\rho=\sum_{i=1}^{N}\,\rho_{0,i}(x)\,\delta(F_{i}),
\end{equation}
where $\rho_{0,i},(i=1,2,\cdots,N)$ are functions defined on layers
$F_{i}=0,(i=1,2,\cdots,N)$. Here the layers are parallel 2-surfaces
in ${\mathbb{R}}^{3}$ having the same normal vectors $\hat{n}$.
Examples are (a) planar multi layers $F_{i}=z-a_{i}=0$ (b) spherical
co-centrical layers $F_{i}=r-a_{i}$ (c) cylindrical co-centrical
layers $F_{i}=r-a_{i}=0$, etc. In all these cases the jump conditions
across the surfaces are given by 
\begin{equation}
\sqrt{\gamma}\,\left(\frac{\partial R_{i+1}}{\partial n}-\frac{\partial R_{i}}{\partial n}\right)+\frac{1}{2}R_{i}^{3}\,\rho_{0,i}=0~~~~\mbox{at \ensuremath{F_{i}=0}},~~i=1,2,\cdots,N,
\end{equation}
where $\gamma$ is the determinant of metric on the two surfaces,
$\frac{\partial R}{\partial n}$ is the derivative along the normal
direction.

Finally let us comment on the scalar field sector of the full theory we have used here. From the vantage point of theory, scalar fields show up in many different settings, yet from the experimental point of view besides the Higgs field, no fundamental scalar field has yet been detected. This of course should not deter one to consider gravity theories that have scalar fields as they could be relevant in the early universe or in the strong field regime of gravity pertaining compact objects. Including the inflaton field that is employed to explain in the initial inflation phase of the universe, there are several we can mention which are consistent with observations. For example,  string-loop modification of the low-energy couplings of the dilaton may provide a mechanism for fixing the vacuum  expectation value of a massless dilaton in a way which is naturally compatible with the existing
 experimental data \cite{dam}. The string expansion involves massless fields other than gravitation, the most relevant being the
 dilaton \cite{Boul}. In a string model there have been some discussions in which the dilaton field changes drastically the
 dynamical properties of the system \cite{Gibb}. A wide class of scalar-tensor theories can pass the present solar-system tests \cite{fujii}
 and still exhibit large, strong field-induced observable deviations in systems involving neutron stars \cite{Dam2}.

\section*{Appendix: Field Equations from an Action}

Let us obtain the field equations (\ref{eq:EoM_gravity_p=00003D0}-\ref{eq:EoM_dilaton_p=00003D0})
from an action. For this, first consider the Einstein-Maxwell-dilaton
action of the form 
\begin{equation}
S=\int d^{4}x\,\sqrt{-g}\left[R-\frac{1}{2}\partial_{\mu}\phi\partial^{\mu}\phi-\frac{1}{4}e^{\gamma\phi}F^{2}\right],
\end{equation}
which is widely studied in the literature; see for example, \citep{Gibbons_et_al-PRL,Gibbons_et_al-PRD}
or \citep{Charmousis_et_al} where the Lagrangian density is augmented
with the Liouville potential $V\left(\phi\right)=2\Lambda e^{-\delta\phi}$.
The field equations of this action are 
\begin{align}
G_{\mu\nu} & =\frac{1}{2}T_{\mu\nu}^{\phi}+\frac{1}{2}e^{\gamma\phi}T_{\mu\nu}^{M},\\
\nabla_{\mu}\left(e^{\gamma\phi}F^{\mu\nu}\right) & =0,\\
\square\phi & =\frac{\gamma}{4}e^{\gamma\phi}F^{2}.
\end{align}
On the other hand, the Einstein-Maxwell action with a constraint on
the norm of the electromagnetic four-potential vector $A^{\mu}$ of
the form 
\begin{equation}
S=\int d^{4}x\sqrt{-g}\left[R-2\Lambda-\frac{c_{1}}{2}F^{2}+\zeta\left(u^{\mu}u_{\mu}+1\right)\right],
\end{equation}
was studied in the literature in the context of Einstein-aether theories
(see for example, \citep{Gurses_Senturk-Godel} and the references
therein) with the field equations 
\begin{align}
G_{\mu\nu}+\Lambda g_{\mu\nu}= & c_{1}T_{\mu\nu}^{M}-\zeta u_{\mu}u_{\nu},\\
\nabla_{\mu}F^{\mu\nu}= & -\frac{\zeta}{c_{1}}u^{\nu},\\
u^{\mu}u_{\mu}= & -1.
\end{align}
The action 
\begin{equation}
S=\int d^{4}x\sqrt{-g}\left[R-\frac{1}{4}e^{-\phi}F^{2}+\zeta\left(u_{\mu}u^{\mu}+e^{\phi}\right)\right],
\end{equation}
where the kinetic term for the scalar field is removed, yields the
field equations 
\begin{align}
G_{\mu\nu} & =\frac{1}{2}e^{-\phi}T_{\mu\nu}^{{\rm M}}-\zeta u_{\mu}u_{\nu},\\
\nabla_{\mu}\left(e^{-\phi}F^{\mu\nu}\right) & =-2\zeta u^{\nu},\\
\frac{1}{4}e^{-\phi}F^{2}+\zeta e^{\phi} & =0,\\
u_{\mu}u^{\mu} & =-e^{\phi}.
\end{align}
The equation coming from the variation of the scalar field yields
the Lagrange multiplier $\zeta$ as 
\begin{equation}
\zeta=-\frac{1}{4}e^{-2\phi}F^{2}.
\end{equation}
Then, the electromagnetic field equation becomes 
\begin{equation}
\nabla_{\mu}\left(e^{-\phi}F^{\mu\nu}\right)-\frac{1}{2}e^{-2\phi}F^{2}u^{\nu}=0.
\end{equation}
At this point, to have a match with (\ref{eq:EoM_electromagnetic_p=00003D0}),
$\rho$ needs to be 
\begin{equation}
\rho=\frac{1}{4}e^{-\phi}F^{2},
\end{equation}
which reduces (\ref{eq:EoM_dilaton_p=00003D0}) to 
\begin{equation}
\square\phi=0.
\end{equation}
At this point, we must emphasize that the variational principle yields
a special case of the equations of motion that we provided in (\ref{eq:EoM_gravity_p=00003D0}-\ref{eq:EoM_dilaton_p=00003D0}).

\end{document}